\documentclass[12pt,preprint]{aastex}

\usepackage{lscape}
\usepackage{textcomp}
\usepackage{graphicx}
\usepackage{epsfig}

\shorttitle{Infall evidence in EGOs} \shortauthors{Xi Chen et al.}

\begin{document}


\title {A Search for Infall Evidence in EGOs I: the Northern Sample
}
\author
 {Xi ~Chen\altaffilmark{1}, Zhi-Qiang ~Shen\altaffilmark{1}, Jing-jing ~Li\altaffilmark{2}, Ye Xu\altaffilmark{2}, and Jin-Hua ~He\altaffilmark{3}}

\altaffiltext{1} {Key Laboratory for Research in Galaxies and
Cosmology, Shanghai Astronomical Observatory, Chinese Academy of
Sciences, Shanghai 200030, China}

\altaffiltext{2} {Purple Mountain Observatory, Chinese Academy of
Sciences, Nanjing 210008, China }

\altaffiltext{3} {National Astronomical Observatories/Yunnan
Observatory, Chinese Academy of Sciences, Kunming 650011, China}


\label{firstpage}


\begin{abstract}

We report the first systematic survey of molecular lines (including HCO$^{+}$
(1--0) and $^{12}$CO, $^{13}$CO, C$^{18}$O (1-0) lines at 3 mm band) towards a new
sample of 88 massive young stellar object (MYSO) candidates associated with
ongoing outflows (known as extended green objects or EGOs)
identified from the Spitzer GLIMPSE survey in the northern
hemisphere with the PMO-13.7 m radio telescope.
By analyzing the asymmetries of the optically thick line HCO$^{+}$
for 69 of 72 EGOs with HCO$^{+}$ detection,
we found 29 sources with ``blue asymmetric profiles'' and 19 sources
with ``red asymmetric profiles''. This results in a blue excess of 0.14,
seen as a signature of collapsing cores in the observed EGO
sample. We found that the sources not associated with IRDCs
show a higher blue excess (0.41) than those associated with
IRDCs (-0.08), the ``possible'' outflow candidates show a higher blue excess
(0.29) than ``likely'' outflow candidates (0.05). A higher
blue excess (0.19) and a lower blue excess (0.07) were also
measured in UC H{\sc ii} regions and 6.7 GHz class II methanol maser
sources, respectively. These suggest the relatively small blue
excess (0.14) in our full sample due to that the observed EGOs are
mostly dominated by outflows and at an
earlier evolutionary phase associated with IRDCs and 6.7 GHz
methanol masers. The physical properties of clouds surrounding EGOs
derived from CO lines are similar to those of massive clumps
wherein the massive star forming cores associated with EGOs possibly embedded.
The infall velocities and mass infall rates
derived for 20 infall candidates are also consistent with the
typical values found in MYSOs. Thus our observations further support
the speculation of Cyganowski et al. (2008) that EGOs trace a
population with ongoing outflow activity and active rapid accretion
stage of massive protostellar evolution from a statistical view,
although there maybe have limitations due to
single-pointing survey with a large beam.

\end{abstract}


\section{Introduction}

There is at present no generally accepted evolutionary scheme for
massive star formation, in contrast to the detailed framework for
the early evolution of low mass stars. Difficulties in the study
of massive star formation lie in the rarity of well studied
sources, due to their short evolution timescale, large distance,
high extinction, and complex star-forming environment within clusters,
typically embedded in dense cores in giant molecular clouds.
Identifying the objects in the early stage of massive star
formation -- massive young stellar objects (MYSOs) is an important
step in attempting to understand massive star formation and its
effect on the evolution of molecular clouds and subsequent star
formation therein.

The IRAS and Midcourse Space Experiment (MSX)
Point Source Catalogs were widely used in previous identifications
of MYSOs (e.g. Sridharan et al. 2002; Urquhart et al. 2008).
However these IR-selected samples are limited by poor resolutions
of IRAS and MSX, thus resulting in maybe still comprising multiple
object emissions within beam size for each candidate.
Recently \emph{Spitzer} Galactic Legacy Infrared Mid-Plane Survey
Extraordinaire (GLIMPSE) using the InfraRed Array Camera (IRAC)
with a resolution less than 2$''$ provides a higher-resolution
dataset to compile new, less-confused samples of IR-selected MYSO candidates.
Cyganowski et al. (2008) have suggested that the 4.5~$\mu$m \emph{Spitzer} IRAC band
offers a new promising approach for identifying MYSOs
with outflows. The strong, extended emission in this band
is usually thought to be produced by shock-excited molecular H$_2$ and CO
in protostellar outflows (e.g. Noriega-Crespo et al. 2004; Reach et al.
2006; Smith et al. 2006; Davis et al. 2007). Such extended 4.5~$\mu$m emission features are
known as ``extended green objects'' (EGOs; Cyganowski et al. 2008) or
``green fuzzies'' (Chambers et al. 2009) for the common color-coding
of the 4.5~$\mu$m band as green in IRAC three-color images.
Cyganowski et al. (2008) have cataloged over 300 EGOs from
GLIMPSE I survey, and they divided cataloged EGOs
into ``likely'' and ``possible'' outflow candidates based
primarily on the angular extent and morphology of the 4.5~$\mu$m
emission. Most EGOs are associated with infrared dark clouds (IRDCs),
and many with known class II 6.7 GHz methanol masers.
Class II 6.7 GHz methanol masers, radiatively pumped by IR emission
from warm dust (Cragg et al. 1992, 2005), are only found to be
associated with MYSOs (e.g. Minier et al. 2003, Xu et al. 2008).
Recent studies have shown that IRDCs are cold ($T<25$ K),
dense (n(H$_{2})$$>$10$^5$ cm$^{-3}$) clouds of molecular gas and dust
(e.g. Egan et al. 1998; Carey et al. 1998, 2000; Simon et al. 2006a, 2006b),
and strong mm or sub-mm dust emissions have been detected in the
cores within them (Beuther et al. 2005; Rathborne et al. 2005, 2006, 2008),
pinpointing IRDCs as sites of the earliest stages of massive star formation.
Though the extended 4.5 $\mu$m emission is also seen towards nearby low-mass outflows
(e.g. Noriega-Crespo et al. 2004),
the close association between EGOs and MYSO tracers (6.7 GHz methanol masers
and IRDCs) provides strong evidence that extended 4.5 $\mu$m emission
from EGOs indeed traces the outflow regions from MYSOs. The cataloged EGOs by
Cyganowski et al. (2008) provide the largest and newest working sample
currently available for massive star formation study.

The dynamics processes in massive star formation regions (MSFRs)
are more complex than in the regions that form low mass stars.
Three main competing concepts of massive star formation have
been discussed in the recent literature, each of which may occur
in nature, depending on the initial and environmental conditions
for the parent molecular clouds (see the recent review of Zinnecker \& Yorke 2007
and references therein). One concept is that stellar collisions
and mergers in very dense systems (Bonnell et al. 1998). The second
is a scaled up version of low-mass star formation (Shu et al. 1987),
via monolithic collapse in isolated cores,
and accompanied by outflow (e.g. Yorke \& Sonnhalter
2002). And the third is competitive accretion in a protocluster environment,
which would also be associated with both large-scale infall and outflows from
individual centre accreting MYSOs (e.g. Bonnell \& Bate 2006).
Moreover, the capture dynamics process by a disk or envelope may also happen
frequently based on the observational evidence of the high frequency
of binary and multiple systems in MSFRs and should not be neglected
in understanding massive star formation (e.g. Moeckel \& Bally 2007a,
2007b). Among these complex dynamics, the role and physics of
accretion are central to understand the massive star formation, yet
remain poorly understood (Beuther et al. 2007a; Zinnecker \& Yorke 2007).
One of the primary questions is that we can not yet
have the resolving power to detect the accretion disk (with a scale
of a few 1000 AU) directly with current instruments, and thus can not answer to
where accretion has been occurred, in an isolated core (monolithic collapse)
or in a protocluster environment (competitive accretion). Over recent years,
a lot of observational evidence showing collimated and energetic outflows in MSFRs
(e.g. Beuther et al. 2002; Xu et al. 2006) indirectly infer
the existence of accretion disk under the assumption that the
high-mass accretion disk drives these outflows via magnetocentrifugal
acceleration. However these outflow evidences still do not give the answer
to where accretion is from an isolated protostar or from a protocluster,
and both monolithic collapse and competitive accretion will be accompanied by
the large scale outflow. Another indirectly tracer for accretion process
in MSFRs is the large scale infall. Even though the large scale infall
is also predicted during massive star formation process by both monolithic
collapse and competitive accretion models, Bonnell \& Bate (2006) have
suggested under the competitive accretion model that infalling signatures
are likely to be confused by the large tangential velocities and the velocity
dispersion presented in the complex cluster environments. From this point of view,
it seems that the large scale infall might be a clue to the
massive star formation.

Millimeter molecular spectral line (e.g. HCN, HCO$^{+}$, CS)
observations have been widely used to search for evidence of
infall in low-mass star forming regions (e.g. Zhou et al. 1994;
Mardones et al. 1997; Lee et al. 1999; Gregersen et al. 1997;
2000) and high-mass star forming regions (e.g. Wu \& Evans 2003,
Fuller et al. 2005; Klaassen \& Wilson 2007; Wu et al. 2007). The
infall motion can be identified from molecular spectral line
asymmetry: an optically thick line (e.g. HCO$^{+}$, CS) in
collapsing core shows a blue asymmetric profile (hereafter named
``blue profile''), which is a combination of double peaks with a
brighter blue peak or a skewed single blue peak, while an
optically thin line shows a peak at the self-absorption dip of
optically thick line. This asymmetry of optically thick line
arises as the blue shifted emission from the approaching warm gas
on the far side of the centre of contraction undergoes less
extinction than the emission from the red shifted, receding,
nearside, material, given that the excitation temperature of the
molecules increases towards the centre of the region (Zhou
1992). Surely outflow and rotation could also produce a blue
profile along a particular line of sight (LOS) to a source, but
infall motion is the only process that would produce consistently
the blue profile. Especially in a large sample of sources the
presence of infall should show a statistically significant excess of blue
profiles compared to red asymmetric profiles (``red profiles'' hereafter), while
an unbiased sample of sources dominated by outflows or rotations
would be expected to show an excess of red profiles or
equal distribution of blue and red profiles, respectively.

An excess of blue profile is now well established
towards low-mass star forming regions as strong evidence
for infall in these regions (e.g. Mardones et al. 1997; Gregersen
et al. 2000; Evans 2003). Recently the infall signatures have also
been found in some molecular line surveys of MSFRs (e.g. Wu \& Evans 2003;
Fuller et al. 2005; Wyrowski et a. 2006; Wu et al. 2007; Klaassen \&
Wilson 2007; Sun \& Gao 2009). The infall signatures also suggest
that the massive star may form by the isolated monolithic collapse mode
as the competitive accretion in a protocluster environment would not
produce the predicted infall signatures (Bonnell \& Bate 2006).
However, some recent surveys also show approximately equal numbers
of red and blue profiles (e.g. Purcell et al. 2006; Szymczak
et al. 2007). Thus further searching for evidence of infall in a
relatively large sample of MYSOs such as EGOs would be important to
enhance our understanding of massive star forming process.
Actually based on the mid-IR colors of EGOs, Cyganowski et al. (2008)
have suggested that most EGOs fall in the
region of color-color space occupied by the youngest MYSOs,
surrounded by substantial accreting envelopes (see Figure 13 in
their work). Therefore EGOs should trace such a population with ongoing
outflow activity and actively rapid accretion stage of massive
protostellar evolution. In this paper, we report the first
systematic 3 mm spectral line survey (including optically thick line
HCO$^{+}$ and optically thin line C$^{18}$O) towards the EGO
sample in the northern hemisphere with the PMO-13.7 m radio telescope.
The observations are described in $\S$ 2; results and analysis are given
in $\S$ 3; and discussions are presented in $\S$ 4, followed by conclusion in
$\S$ 5.

\section{Observations}

Single-point observations were
carried out towards 88 EGOs with $\delta>$-20$^{\circ}$ compiled
from EGO catalog (Cyganowski et al. 2008) using the
13.7 m telescope of Purple Mountain Observatory (PMO) in Delingha,
China. We did not include those nearby EGOs with position
separation of less than 40$''$ (corresponding to half the beam size
of the telescope at HCO$^{+}$ observations; see below) from the target sources. The sample includes 50
``likely'' MYSO outflow candidate EGOs selected from Tables 1, 2
and 5 of Cyganowski et al. (2008) and 38 ``possible'' MYSO outflow
candidate EGOs selected from Tables 3 and 4 of Cyganowski et al.
(2008). Table 1 lists the sample source parameters including
name, position, association with IRDC, 6.7 GHz class II
methanol maser, Ultracompact (UC) H{\sc ii} region and 1.1 mm
continuum source. We consider a class II methanol maser,
UC H{\sc ii} region or 1.1 mm continuum source to be
associated with an EGO if the separations between them
and the EGO is less than 30$''$. The positional
accuracies of the {\em Spitzer} GLIMPSE point source catalog,
and 6.7 GHz methanol maser catalogs (Szymczak et al. 2007;
Cyganowski et al. 2008, 2009; Caswell 2009, Xu et al. 2009) and
UC H{\sc ii} region catalogs (Wood \& Churchwell 1989; Becker et al. 1994;
Kurtz et al. 1994; Walsh et al. 1998; Forster \& Caswell 2000) used in our analysis,
are usually better than 1$''$. And the positional uncertainty of 1.1 mm continuum source presented
by BOLOCAM Glactic Plane Survey (GPS) is also at the order of several
arcseconds (Rosolowsky et al. 2009). However, EGOs are extended
objects with angular extents between a few to $>$30$''$ (Cyganowski et al. 2008),
and the emission from 6.7 GHz methanol maser, UC H{\sc ii} or 1.1 mm continuum
regions within 30$''$ can be encompassed by the \textbf{PMO} telescope
beam, even if they may not be physically associated.
The observed lines include the optically thick
line HCO$^{+}$ ($1-0$) and optically thin line C$^{18}$O (1--0),
as well as $^{12}$CO (1--0) and $^{13}$CO (1--0) lines.
The lines, observing date, rest frequency, beam size, main-beam
efficiency, and velocity resolution are summarized in Table 2.

We performed HCO$^{+}$ (1-0) line observations in
Jan 9-23, 2009. The half-power beamwidth was about 80$''$ (Table 2).
A cooled SIS receiver working in the 80-115 GHz band was employed.
The system temperature was around 190-290 K and typical atmospheric
absorption $\tau$ was about 0.10-0.15, depending on the weather
conditions. The backend was a Fast Fourier Transform Spectrometer
(FFTS) of 16384 channels with bandwidth of 1 GHz and an equivalent
velocity resolution of about 0.21 km s$^{-1}$. The observations
were performed in a position-switching mode with off positions
offset 10$'$ in the direction away from the galactic plane
(no emission was found in each off position).
The pointing rms was better than 10$''$.
The standard chopper wheel calibration was used during
the observations to get the antenna temperature, T$_{A}^{*}$, which
has been corrected for atmospheric absorption. The main beam
efficiency $\eta_{mb}$ is 0.50. Then the line intensity was reported
in terms of the corrected main beam temperature T$_{MB}$ =
T$_{A}^{*}$/$\eta_{mb}$, and the uncertainty is about 20\%. For the
first step, we performed HCO$^{+}$ line observations with an
on-source integration time of 10 minutes towards full sample,
yielding a 1$\sigma$ noise level of T$_{A}^{*}$ about 0.09 K. Then
additional 5-15 minute on-source time (depending on the HCO$^{+}$
intensity of each source) observations were continued towards 72
sources with HCO$^{+}$ emission detected in initial observations
to improve the spectral line profiles. This finally yielded a typical rms
noise level of 0.05-0.08 K (T$_{A}^{*}$) for the HCO$^{+}$ line.

In order to obtain the line center velocity information, we
carried out the optically thin line C$^{18}$O (1--0) observations towards
these 72 EGOs with HCO$^{+}$ detection. All the three CO ($^{12}$CO,
$^{13}$CO and C$^{18}$O (1--0)) lines were observed simultaneously. The CO observations were
carried out in Feb 12-17, 2009. The beam size was about 65$''$ at 110
GHz. The same receiver was used. The system temperature was about
200-300 K, and typical atmospheric absorption $\tau$ was about 0.1-0.2.
The backends were three Acousto-Optical Spectrometers (AOS) of 1024 channels
with bandwidth of 42.7, 43.2, and 145.4 MHz and the corresponding velocity resolution of about
0.11, 0.11, and 0.37 km s$^{-1}$, for C$^{18}$O, $^{13}$CO and $^{12}$CO,
respectively. The observations were also performed in a position-switching
mode, but the off positions were found in 1$^\circ$ offset from targeted
positions to avoid the background emission, and
were checked with no emission. The on-source time is about 5-20 minutes
for each source, which led to the typical rms noise of 0.04-0.11 K
(T$_{A}^{*}$) for C$^{18}$O line. The main beam efficiency
$\eta_{mb}$ is about 0.60 in this observing mode.

The spectral data were reduced and analyzed with the GILDAS/CLASS
package. In the process, the baseline subtraction have been
performed on the spectra.

\section{Results and Analysis}

\subsection{Spectral-line Detections}

In our observations of 88 EGOs (Table 1), HCO$^{+}$ (1--0) emission was
detected in 72 sources with a detection limit of 0.50 K ($3\sigma$) at
T$_{MB}$ scale, giving a detection rate of about 80\%. Such a high
detection rate demonstrates the presence of rich molecular gas in EGOs.
Interestingly, none of those 16 EGOs without detected HCO$^{+}$ is associated
with the known UC H{\sc ii} region, suggesting that the non-detected sources
may be at the very earlier evolutionary stage (possibly prior to the
phase of formation of a dense core), so that the
gas density is still below the critical density of HCO$^{+}$
(1--0) of 5$\times$10$^{4}$ cm$^{-3}$.

The HCO$^{+}$ line parameters of all the 88 sources observed
are given in Table 3. The HCO$^{+}$ line profiles of the most
sources show blue or red asymmetric line profiles. We adopted
the same analysis method of Purcell et al. (2006) to parameterize
the HCO$^{+}$ line for sources with asymmetric line profiles
(double-peaked and skewed profiles): the absorption dip was
blanked so that a single-Gaussian fitting routine was constrained
only by the sides of the line, and the distinct residual line wings
were subtracted during the fitting routine. The Gaussian fitting
parameters after subtracting Gaussian fits to any line wings,
the integrated intensity, $\int$T$_{MB}dV$, the velocity at peak,
V$_{LSR}$, the line width, $\Delta$V and the peak main beam
temperature, T$_{MB}$, are given in columns (2), (3), (4) and
(5) of Table 3, respectively. We also list the actual observed T$_{MB}$
and V$_{LSR}$ estimated directly from the HCO$^{+}$ spectrum for
each double-peaked and skewed profile in columns (6) and (7)
of Table 3, for analyzing the line asymmetry compared to the
optically thin line (see \S 3.2). For non-asymmetric line
profile sources, we only performed single Gaussian
fits to them, and the actual observed T$_{MB}$ and V$_{LSR}$ adopt
the Gaussian fit values in columns (5) and (3) to analyze
the HCO$^{+}$ line asymmetries.

Recent surveys of HCO$^{+}$ at different transitions, e.g. at $1-0$
(Purcell et al. 2006) and $4-3$ (Klaassen \& Wilson 2007) have shown
that HCO$^{+}$ spectra can provide evidence of outflows in MSFRs.
In our observations, 30 sources out of
72 with HCO$^{+}$ line detection show residual
line wings distinct from the main line, suggesting the possible existence of
molecular outflows among these sources, and thus the outflow detection rate
would be $\sim40\%$ with HCO$^{+}$. These line wings were fitted
by a single broad Gaussian and then were subtracted from the detected line
before performing analysis for the main line properties described as above.
The line wing parameters from Gaussian fits are listed in Table 4.
We categorized the line wings to three groups based on their shapes:
``D'' -- double wings, means the line wing emission from both
blue and red wings; ``B'' -- blue wing, and ``R'' -- red wing, denote
the line wing mainly appearing in blue and red part, respectively.
Of them, 21 sources show broad double wing emission, 3 and 6 sources only
show blue line wing and red line wing emission respectively.

Comparison with the HCO$^{+}$ ($1-0$) survey towards the southern
methanol masers using the Mopra telescope reported by Purcell et
al. (2006), 9 sources overlap between our survey and their survey,
and 3 of them also show broad line wing emissions. We list the main
line and line wing parameters of these sources acquired in our survey
and their survey in Table 5. The observed main line profiles among
these sources in our survey are similar to that in Purcell et al. (2006),
but the line emission intensities in our survey are about 50\% lower
than that presented in Purcell et al. (see main line T$_{MB}$
listed in Table 5). The weaker emission is mainly because the
HCO$^{+}$ emission undergoes strong beam dilution effect in our
observations with a relatively large beam size ($\sim$80$''$)
compared to 35$''$ in Purcell et al. (2006).
This suggests that the EGO emission regions do not fully
fill in the whole beam size of the PMO telescope. All the 3 sources with broad line wings
overlapped in the two surveys also show similar wing shapes (all double wings),
even if the line wing width is significant different
in two sources (G10.34-0.14 and G12.91-0.26; see Table 5).
The similar spectral profiles (including main line and line wing)
of these sources also indicate that EGOs would dominate the observed HCO$^{+}$ line,
although there may be some other sources in addition to EGOs
at present within the telescope beam. But, we could not
exclude absolutely that the observed emission might be dominated by other
sources within the beam.

Asymmetric line profiles may be interpreted by
either multiple emitting regions along the same LOS, or a single
emitting region with cold absorbing gas intervening. The line center
velocity determined from optically thin line is needed to distinguish
two cases. Although the optically thin line H$^{13}$CO$^{+}$ is the
best tracer for the line center of HCO$^{+}$ line, the expected
H$^{13}$CO$^{+}$ emission in EGOs is too weak to be easily detected.
Instead, we used the optically thin line C$^{18}$O
to estimate the line center velocity. We performed
simultaneous observations of three CO ($^{12}$CO, $^{13}$CO and C$^{18}$O)
lines towards 72 EGOs with detected
HCO$^{+}$ emission as listed in Table 3. All these 72 sources were detected
emission in all three CO lines. We did single-Gaussian fit to each of
three CO lines for each source. The Gaussian fit results of three
CO lines for all 72 sources are summarized in Table 6. All sources
but for G25.27-0.43 show only one C$^{18}$O velocity component with
a single peak near the velocity position of HCO$^{+}$ line.
This suggests that the line center information obtained from C$^{18}$O
line should suffer little confusion from background and foreground emission
along LOS.

However, the $^{12}$CO line properties derived
from a single-Gaussian fit presented here may be not very
accurate as that in some sources $^{12}$CO lines tend to be strong self-absorption
or show multi-velocity components around the velocity position of
HCO$^{+}$ due to confusion from the background and foreground
emission.

\subsection{Blue Profile Identification}

Among the 72 sources with the detected HCO$^{+}$ emission, three sources
(G12.68-0.18, G25.27-0.43, G28.85-0.23) show too complex spectral profiles
of either HCO$^{+}$ or CO lines, thus are ignored in our further
analysis for the line asymmetry. Usually two methods of characterizing
line profiles were used in the literature. When the opacity is high enough and the
line takes on a double-peaked profile, a blue profile caused by
infall motion with velocity $v\thicksim r^{-1/2}$ in a region with
higher excitation temperature ($T_{ex}$) inside requires
$T_{MB}(B)/T_{MB}(R)>1$. Here $r$ is the radius of the
collapsing core, $T_{MB}(B)$ and $T_{MB}(R)$ are the blue and
red peak main beam brightness temperature of the optically thick lines (Zhou et
al. 1993). At lower optical depths the absorption will be less
severe and the line will appear as a skewed peak with a red or
blue shoulder. An alternative definition (Mardones et al. 1997) is
used for these cases as well. A line can be identified as a blue
profile if the peak of the optically thick line is shifted
blueward, with the velocity difference, $\delta v$, between the
peaks of the optically thick line and the optically thin line
greater than a quarter of the line width of the optically thin
line: $\delta v$=($v_{thick}$-$v_{thin}$)/$\Delta
v_{thin}$$<-0.25$. Here $v_{thick}$ is the peak velocity of the
optically thick line; $v_{thin}$ and $\Delta v_{thin}$ denote the
peak velocity and width of the optically thin line. In contrast, a
red profile would have $\delta v>0.25$. The adpoted threshold value 0.25
corresponds to about 5 times the typical rms error in $\delta v$
(see Mardones et al. 1997).

The parameters T$_{MB}$(B)/T$_{MB}$(R), $\delta v$ with their
uncertainties for 69 sources in our analysis sample derived from
above two methods are given in Table 7. For double-peaked profile,
we used the velocity at the brightest peak as the value of $v_{thick}$
to calculate $\delta v$. Following Mardones et al. (1997) we also
adopt $\delta v$ of $\pm$0.25 (also corresponding to 5 times the
typical rms error in $\delta v$ listed in column (2) of Table 7.)
as the threshold to define the line asymmetries. As for brightness
ratio method, we consider an asymmetry to be significant
if the difference (or sum) between T$_{MB}$(B)/T$_{MB}$(R) and its
uncertainty is still larger (or less) than 1 for blue (or red)
double-peaked profile. Finally we identify 29 blue asymmetric profile candidates
(including 20 double-peaked profiles and 9 skewed profiles) and
19 red asymmetric profile candidates (including 9 double-peaked profiles and
10 skewed profiles). The remaining 21 sources do not show
significantly asymmetric profiles. Note that all 20 blue double-peaked
profiles or all 9 red double-peaked profiles can also be defined
as blue or red profiles based on the $\delta v$
method. That is, the blue or red
profiles identified by the brightness ratio method is a subset of the
blue or red profiles identified by the $\delta v$ method.
The profile asymmetries for all the 69 sources
are summarized in column (4) of Table 7.
For seeing clearly the HCO$^{+}$ asymmetry characteristics,
we show the HCO$^{+}$ and C$^{18}$O spectra of 29 blue profile
candidates and 19 red profile candidates in Figures 1
and 2, respectively. The spectra of the remaining 21 sources with
no significantly asymmetric profiles, as well as the 3 sources
with complex HCO$^{+}$ or C$^{18}$O spectra are shown in Figure 3.
We also give the $^{12}$CO and $^{13}$CO spectra of all the 72 sources in
an online figure.

In our observations, the optically thin C$^{18}$O line shows an ideal
Gaussian profile and is fitted by the Gaussian function to obtain
the line center velocity and width. We also simply estimate the
optical depth for C$^{18}$O based on Myers et al. (1983) method:
\begin{equation}\label{1}
\frac{(T_{MB})_{13}}{(T_{MB})_{18}}=\frac{1-exp(-5.5\tau_{18})}{1-exp(-\tau_{18})},
\end{equation}
where (T$_{MB}$)$_{13}$ and (T$_{MB}$)$_{18}$ are the peak main beam
temperature of $^{13}$CO and C$^{18}$O; $\tau_{18}$ is the optical
depth of C$^{18}$O. This method was widely adopted in the analysis
for dark clouds that usually show strong $^{12}$CO line self-absorption
or blended profile due to background emission. Thus this method should
well adapt to our cases as most sources in our observations also show
apparent $^{12}$CO line self-absorption or blended profiles. However
when $(T_{MB})_{13}/(T_{MB})_{18}>5$ appears in our calculations,
the derived optical depth of C$^{18}$O seems unreasonably
small (this would lead to a larger value of excitation
temperature). For these cases (marked by ``b'' in Table 6) we adopted
the typical LTE method to estimate the optical depth using
$^{12}$CO peak temperature (e.g. Sato et al. 1994). The derived
optical depth of C$^{18}$O is listed in column (5) of Table 7. It can be
seen that for all sources the optical depth is less
than 1 with a mean of 0.2. This suggests the C$^{18}$O line
is indeed optically thin in our sample, therefore it can reach the
region near the centre of collapsing core and can be used to
determine the system velocity.

Another question about using
C$^{18}$O to determine the line center velocity comes from whether
HCO$^{+}$ and C$^{18}$O lines are tracing the same material. From
Figures 1, 2 and 3, it can be clearly seen that C$^{18}$O can
accurately trace the line center velocity in all sources except
for G49.27-0.34 as shown in Fig. 3. From its HCO$^{+}$ spectrum,
G49.27-0.34 seems to show a blue profile somewhat with a
self-absorption dip at velocity of 71.5 km s$^{-1}$, while the
line center velocity determined from C$^{18}$O is 67.8 km
s$^{-1}$. This source is considered to be the non-asymmetric
profile based on the $\delta v$ method in our work. It
suggests that C$^{18}$O emission should trace the same material as
HCO$^{+}$ line and thus be good to determine the center velocity
of HCO$^{+}$ line. Moreover some mapping observations
of HCO$^{+}$ ($1-0$) and C$^{18}$O (1-0) lines also provide the
evidence that both lines can trace the same emission region
(e.g. Qi et al. 2003; Wu et al. 2009). Sun \& Gao (2008) also
suggest that $^{13}$CO and C$^{18}$O lines can be used to measure
the line center velocity of HCO$^{+}$ with the same PMO-13.7 m telescope.

Comparison with the high transition HCO$^{+}$ (3--2) spectra
acquired with the James Clerk Maxwell Telescope (JCMT) by Cyganowski et al.
(2009), we surprisingly found that only 3 sources show the same
line profile classification of blue/red or non-asymmetric profiles
in a total of 17 common sources (comparing Table 7 in our work with Table 6 in
Cyganowski et al. 2009). One possibility for such a difference between
two studies is that the spectral resolution of HCO$^{+}$ (3--2)
of $\sim$0.6 km s$^{-1}$ is different from our HCO$^{+}$ (1--0) spectra with
$\sim$0.2 km s$^{-1}$. For checking this, we only consider 6 sources
(G11.92-0.61, G19.36-0.03, G22.04+0.22, G28.28-0.36, G28.83-0.25 and
G35.03+0.35) with clear HCO$^{+}$ (3--2) self-absorption dips. But there
are still 3 sources (G19.36-0.03, G28.28-0.36 and G35.03+0.35) show contrary
profile classifications. It suggests that the spectral resolution difference
is not the major factor for the profile classification difference in the
two studies. Another possibility is a larger beam (80$''$) used
in our HCO$^{+}$ (1--0) observations relative to 19$''$ in the JCMT
HCO$^{+}$ (3--2) observations. However all the 9 overlapped sources between
our survey and Purcell et al. (2006) survey show similar HCO$^{+}$ (1--0) profile
even though the two surveys have different beam sizes (see \S3.1).
And none of all the 5 sources overlapped among the three surveys (HCO$^{+}$ (1--0) by our observation
and Purcell et al. 2006, HCO$^{+}$ (3--2) by Cyganowski et al. 2009)
shows the same profile classifications based on HCO$^{+}$ (1--0) and HCO$^{+}$ (3--2) spectra.
Thus combining these factors, it is most likely that the different transitions of
HCO$^{+}$ may trace different gas circumstances and need different excitation conditions
to shape the corresponding blue or red profiles. This point has also been reported by
Fuller et al. (2005) and Wu et al. (2007) that showed the different line asymmetries
of HCO$^{+}$ at transitions of (1--0) and (3--2) in some observed sources.

\subsection{Blue Profile Excess Quantity}

Most mechanisms that can produce asymmetric line profiles towards
sources, e.g. rotation, outflow, should have approximately
equal numbers of red and blue profiles, except that
infall preferentially produces blue profiles. To
quantify whether blue profile dominates in a sample, Mardones et
al. (1997) defined a quantity E, the blue excess:
$E=(N_{blue}-N_{red})/N_{total}$, where $N_{blue}$ and $N_{red}$
are the number of sources which show blue or red profiles,
respectively, and $N_{total}$ is the total numbers of sample
sources. Since then, the blue excess quantity has been widely adopted
in line asymmetry analysis (e.g. Wu \& Evans 2003, Fuller et al.
2005, Purcell et al. 2006, Wu et al. 2007). To compare with other
surveys we also adopt the blue excess quantity analysis in our
sample. In order to see the blue excess more clearly, we plot
the T$_{MB}$(B)/T$_{MB}$(R) and $\delta v$ of HCO$^{+}$
distributions in Figure 4. The blue excess using the $\delta v$
and brightness ratio methods is (29-19)/69=0.14 and (20-9)/69=0.16, respectively
and with corresponding probability, $P$, of 0.09 and 0.02 that arises by chance.
The probability, $P$ that the blue excess arises by chance is estimated
with the binomial test (see Fuller et al. 2005 and references
therein). We adopted the blue excess value derived from $\delta v$
method because this can identify a relatively large number of sources
with line asymmetry. The statistical results are summarized in
Table 8.

\section{Discussion}

\subsection{Blue Profile Excess in Different Sample}

The blue excess in our EGO sample is similar to that reported in Fuller et al. (2005).
They also found the blue excess of 0.15 with the same HCO$^{+}$
(1--0) line in a sample of 77 MYSOs with distances ranged
from 1 to 10 kpc and associated with 850 $\mu$m emission. However,
these values are distinctly smaller than the results reported in other
surveys. For comparison, Wu \& Evans (2003) found 12 blue
profile candidates and measured a blue excess of 0.29 among a
sample of 28 MYSOs with distances spanned a range of 1 -- 12 kpc
and associated with water masers; Wu et al. (2007) detected
blue profiles in 17 cores and blue excess of 0.29 within a sample
of 46 MYSOs associated with precursors of UC H{\sc ii} and
UC H{\sc ii} regions and with distances ranged from 0.5 to 10 kpc.
Whereas the blue excess of 0.14 found in our survey is slightly
higher than that reported by two recent major HCO$^{+}$ (1--0)
studies of 6.7 GHz methanol maser sources by Purcell et al. (2006)
and Szymczak et al. (2007). They found that approximately equal numbers
of red and blue profiles, i.e. no infall signatures in
their survey. For understanding the blue excess obtained in our
survey and comparing with other surveys in detail, we divided
our sources into some sub-samples as follows.

\subsubsection{Distance Effect}

Fuller et al. (2005) suggested that the lower blue excess value found in
their work is most likely because the telescope beam may contain
emission from material not intimately associated with MYSOs for
those at larger distances. To limit this possibility, they
carried out the analysis for sources with distances of less than 8
kpc, and found blue excess of HCO$^{+}$ (1--0) line to be 0.28. We
performed such an analysis to minimize the distance effect. We
used new galactic rotation curve (Reid et al. 2009) to calculate
the kinematic distances, assuming the galactic
constants, R$_{\odot}=$ 8.4 kpc and $\Theta_{\odot}=$ 254 km
s$^{-1}$. The system velocity was determined from Gaussian fits
to the C$^{18}$O line. Most sources lie inside the solar circle
and thus may have a near/far distance ambiguity, the near kinematic
distances were adopted for these sources in our analysis, and listed
in column (6) of Table 7. However the distances
for four sources (G49.07-0.33, G49.27-0.32, G49.27-0.34 and G59.79+0.63) can not
be derived from the galactic rotation curve, 5 kpc was assumed.
The distances of most sources span a range of 2-6
kpc except for two sources (G12.02-0.21 and G49.42+0.33) with
distances of larger than 12 kpc. For comparison, a distance of 4
kpc was adopted as a limit, which is compared to 8 kpc adopted in
Fuller et al. (2005) with a telescope beam of $\sim30\arcsec$. We
formed a sub-sample of 31 objects with distances of less than 4
kpc and have looked at the blue excess for these objects alone.
For this distance limited sub-sample, 12 blue profile candidates
and 10 red profile candidates were found -- thus an even smaller blue
excess of only 0.06 with a probability of 0.58 appeared (see Table 8).
It suggests that the far
distance effect did not play an important role in understanding
the relatively small blue excess in our sample. However,
the adopted near kinematic distance may not be very reliable. If
the red profile candidates have larger distances i.e. at far
kinematic distances, a higher blue excess would be
expected. Another possible explanation for relatively small
blue excess even after considering the distance effect is the small EGO
size with typical less than 30\arcsec\ seen from their IRAC
images, which makes the telescope beam still encompass the emission
from material not intimately associated with MYSOs even at smaller
distances.

\subsubsection{``Likely'' and ''Possible'' Outflow Candidates}

According to Cyganowski et al. (2008), the 69 sample sources used in
our line asymmetry analysis include 41 ``likely'' MYSO outflow candidates
and 28 ``possible'' MYSO outflow candidates. We performed the blue
excess analysis for the sub-samples of ``likely'' and ``possible'' outflows,
respectively. We found 14 blue profiles and 12 red profiles of
41 ``likely'' outflow candidates, and 15 blue profiles and 7 red
profiles of 28 ``possible'' outflow candidates. Thus
the blue excess seen in ``likely'' and ``possible'' outflow
candidates is 0.05 and 0.29, respectively with corresponding probability of
0.42 and 0.07 (see Table 8). The blue excess in
``likely'' sub-sample is distinctly smaller than that in
``possible'' sub-sample. The appearance of very low blue excess in ``likely''
sub-sample is most likely because these sources are dominated by outflows.
The distinct extended 4.5 $\mu$m emission around
these sources identified from IRAC images suggests that they
should be associated with strong and distinct outflows, thus were
classified into the ``likely'' MYSO outflow candidates by
Cyganowski et al. (2008). The significant outflows may be shaping
more red profiles which results in weakening the blue excess.
On the other hand the relatively high blue excess
of 0.29 found in ``possible'' outflow sub-sample similar to
the typical value found in other surveys (e.g. Wu \& Evans 2003, Wu et al.
2007) suggests that the outflows may be weak and unobvious in
``possible'' sub-sample compared to ``likely'' sub-sample, consistent with
the outcomes seen in IRAC images (Cyganowski et al. 2008).

We also found that 18 of 42 ``likely'' outflow candidates, and 12 of
30 ``possible'' outflow candidates show broad HCO$^{+}$ line wing emissions, suggesting
the possible existence of outflow among these sources (note that additional 1 ``likely'' and 2
``possible'' outflow candidates that were not used in the line asymmetry analysis
are included in this statistics). But the detection rates of
outflow traced by HCO$^{+}$ line wings are similar in both ``likely'' outflow sub-sample (18/42=43\%)
and ``possible'' outflow sub-sample (12/30=40\%). Thus it seems that the HCO$^{+}$ line wing
emissions are not sensitive to the outflow classifications (i.e. ``likely'' and ``possible'')
seen from the IRAC images.

\subsubsection{IRDCs and non-IRDCs}

We also divided our sample into two sub-samples depending on
their association with IRDCs. Usually IRDCs are believed
to be potential sites of massive star formation and represent
early stage of massive star formation. There are 12 blue profiles
and 15 red profiles, and thus a negative blue excess of -0.08 with
a chance probability of 0.65 existed among 37 IRDC sources, whereas
17 blue profiles and 4 red profiles, and a blue excess of 0.41 with
a small chance probability of 0.004 were found among 32 non-IRDC sources.
The statistical results are also given in Table 8.

The blue excess
found in IRDC sub-sample is much smaller than that in non-IRDC sub-sample.
One explanation for different blue excess between them is that
the infall signatures may be different at different evolutionary
stages of massive star formation. Actually the relation between
blue profile excess and evolutionary phase has been \textbf{studied} in low-
and high-mass star forming regions (e.g. Mardones et al. 1997, Wu
et al. 2007). In low-mass cores the blue excess was found to be
0.30, 0.31, and 0.31 for --I, 0, and I core samples, respectively (Evans 2003
and references therein). There seems to be no significant
difference among different evolutionary phases of low-mass star
formation. In massive star forming cores, UC H{\sc ii} regions
show a higher blue excess (0.58) than UC H{\sc ii} precursors
(0.17), indicating that material is still accreted after the onset
of the UC H{\sc ii} phase and there has a higher blue excess at
the stage with hot cores (Wu et al. 2007). This may point to
fundamental difference between low- and high-mass star forming
conditions. In our observations, the very low or negative blue excess
detected in the early evolutionary stage sources associated
with IRDCs can be explained as follows: (1) the molecular gas in IRDCs
surrounding EGOs (i.e. earlier stage sources) may not be
adequately thermalized to show the blue excess; (2)
The amount of dense cool gas is larger toward younger objects.
Outflows of dense molecular gas may be more active around IRDC
sources, shaping more red profiles. One evidence for this is that most
(26/41=63\%) ``likely'' outflow candidates are associated with
IRDCs. There is also a slightly higher detection rate of the outflow traced by
HCO$^{+}$ line wing of 18/38=47\% in IRDC sub-sample than
that of 12/34=35\% in non-IRDC sub-sample (here in both sub-samples, we
include additional 1 IRDC and 2 non-IRDC sources that were not used
in the line asymmetry analysis). These suggest that
the outflows are indeed more active around IRDC sources at very
early evolutionary stage of massive star formation, and thus shaping
more red profiles and resulting in lower blue excess. Comparison with
IRDCs, the higher blue excess found in non-IRDCs may be due
to that these sources are at a relatively late stage of massive
star formation. However, we have no justification to believe that
sources not associated with IRDCs must be located at
the late evolutionary stage. The visibility of an IRDC is dependent
on the strength of the mid-infrared background emission particularly
at 8 $\mu$m (Cyganowski et al. 2008). If there is no or weak 8
$\mu$m emission in a particular region, non-IRDC may be visible,
even if dense molecular gas and very young MYSOs are present.

Some other factors could not be neglected to understand the different
blue excess between IRDC and non-IRDC sub-samples. For example,
molecular material in the IRDC but not directly physically associated
with the EGO may contribute significantly to the emission within the large
beam. Dynamical interactions between outflows and surrounding
molecular material in IRDC sub-sample may be very different from those
in non-IRDC sub-sample. Both above possibilities would affect the line profiles
and result in different observed blue excess. Moreover, the lower
blue excess in IRDC sub-sample seems to be also explained if EGO associated
with IRDC is in the cluster environment and wherein the competitive
accretion is occurred. As Bonnell \& Bate (2006) suggested, the larger
tangential velocities and the velocity dispersion presented in the cluster
environments would reduce the infall signatures under the competitive
accretion. Obviously, our current single-pointing observations with a large
beam can not give answers to any possibilities described as above.
Higher-angular resolution data are needed to clarify these possibilities.

\subsubsection{6.7 GHz methanol masers and UC H{\sc ii} Regions}

In addition, our sample is also associated with other
astrophysical objects, e.g. UC H{\sc ii} regions and 6.7 GHz
class II methanol masers. We carried out the blue excess analysis
for these sub-samples with the statistical results also listed in Table
8. We found that the blue excess in UC H{\sc ii} regions ($\sim0.19$)
is slightly higher than that in 6.7 GHz methanol maser sources ($\sim0.07$).
The different blue excess in these two sub-samples seems also to
be due to that the blue excess may evolve with the different
massive star formation stage as described in above section.
The 6.7 GHz class II methanol masers are only associated with MSFRs
and also trace an early evolutionary stage, as
evidenced by their association with IRDCs (Ellingsen 2006) and millimeter
and submillimeter dust continuum emission (Pestalozzi et al. 2002; Walsh
et al. 2003). Thus the blue excess could be lower in 6.7 GHz methanol maser sub-sample
due to not adequately thermalized gas in the centre of the core at the earlier
stage. Two recent HCO$^{+}$ surveys of 6.7 GHz methanol maser sources by
Purcell et al. (2006) and Szymczak et al. (2007) that also found no or very
low blue excess in their samples are consistent with our statistical result.

As stated in previous section, Wu et al. (2007) found a higher blue excess
(0.58) in UC H{\sc ii} regions. However the blue excess in UC H{\sc ii}
regions detected in our survey is quite lower. One
possibility is that the sources associated with UC H{\sc ii} regions may
be at a range of evolutionary stages. For clarifying this, we exclude the
possible younger sources associated with both UC H{\sc ii} region and
methanol maser in our analysis. Interestingly, we found 5 blue profiles and only 1
red profile among a total of 6 sources that are only associated with UC H{\sc ii}
region and are believed to be at late evolutionary stage,
thus the blue excess is as high as 0.67. Such a high blue excess is
in agreement with that reported by Wu et al. (2007).

Combining above results from our survey and other surveys, it seems that
the blue excess evolving with different stage may be a genuine
characteristic of massive star formation. However the small samples of
UC H{\sc ii} region (16) and methanol maser (25) and the larger chance
probabilities of blue excess (0.3-0.4) in our surveys (see Table 8) should
be taken into account before drawing any definite conclusions. It
should be noted that not all of the observed northern EGOs have been
searched for UC H{\sc ii} regions or 6.7 GHz methanol masers
(not all fall within the coverage of large blind surveys for these tracers),
thus there may be a bias in the known associations of EGOs with these tracers.

\subsection{Physical Properties of the EGOs}

Cyganowski et al. (2008) argued that GLIMPSE-identified EGOs are only associated with
high-mass star formation based on their association with other
high-mass star formation tracers such as IRDCs and class~II
methanol masers, and that the surface brightness of low-mass
outflows will be too faint to be detected at the sensitivity of
the GLIMPSE survey, even though the extended 4.5 $\mu$m could
also be seen towards nearby low-mass outflows as well
by deep observations (e.g. Noriega-Crespo et al. 2004).
Chen et al. (2009) analyzed the EGOs searched for
class I methanol masers (61 in total) and found that
class~I methanol masers were detected towards about two-thirds
of EGOs, also suggesting that EGOs are associated with outflows from
MYSOs. Moreover comparison with recent published 1.1 mm continuum BOLOCAM GPS
catalog (Rosolowsky et al. 2009), we found that most (65/77=84\%;
see Table 1) EGOs in our observed sample are associated with
mm continuum source within 30$''$ (corresponding to half the PMO beam size).
The detected mm continuum emission, especially in IRDCs, suggests that EGOs
pinpoint the sites of massive star formation at early evolutionary
stage (Rathborne \& Jackson 2006). However, the physical properties
of the EGOs such as gas density and mass etc.
are still unclear.

To understand their physical properties, we
calculated the parameters with C$^{18}$O line following the
typical LTE method (e.g. Sato et al. 1994). The physical parameters
derived from C$^{18}$O line are also given in Table 7.
The typical size of EGO of a few to 30$''$ seen
from IRAC images is smaller than the beam size of the telescope.
Thus it is very likely that the C$^{18}$O emitting region is too
small to fully fill in the whole telescope beam. We determined the beam
filling factor, $f$, for each source
with the same method adopted in Klaassen \& Wilson (2007)
that assumed a consistent ambient temperature for all the
observed sources (Eq. 2 in their work). In our observed
EGOs, one source G34.26+0.15 has a similar scale ($\sim$ 60$''$
seen from its IRAC image) to the telescope beam size, we assumed
the beam filling factor is 1 for this source. Under this assumption
and Eq. (2) of Klaassen \& Wilson (2007), an ambient temperature of 30 K
was derived for this source. We adopted this ambient temperature
to all the other observed EGOs. The derived beam filling factor for each
source is listed in the column (7) of Table 7. And then the angular size of
the C$^{18}$O emitting region was estimated from the beam size and beam filling factor.
Linear size (column (8) in Table 7) was calculated from the
angular size and kinematic distance (column (6) in Table 7).
The mean size of the C$^{18}$O emitting region is 0.8 pc (typically
0.5-1 pc). The derived mean column density of C$^{18}$O (column (9)
in Table 7) is $3\times10^{16}$ cm$^{-2}$.
The column density of H$_{2}$ is derived with a moderate
C$^{18}$O abundance of $1.7\times10^{-7}$ (Frerking et al. 1982).
The mean column density of H$_{2}$ (column (10) in Table 7)
is $2\times10^{23}$ cm$^{-2}$, while the mean volume
density (column (11) in Table 7) is $1.5\times10^{5}$ cm$^{-3}$.
The LTE masses based on C$^{18}$O (column (12) in Table 7) range from a few 100 to
several 1000 M$_\odot$ with a mean mass of $2\times10^3$ M$_\odot$
(this mass range is very similar to that for the sample of 6.7
GHz methanol maser sources and UC H{\sc ii} regions reported
by Purcell et al. (2009)). All above physical parameters
are consistent with the characteristics of the massive
clumps wherein massive star forming cores associated
with EGOs possibly embedded (e.g. Beuther et al. 2007). Here we follow
Beuther et al. (2007) to define
the terms clump and core: clump for condensations
associated with cluster formation, and core for molecular
condensations that form single or gravitationally bound multiple
massive protostars. Moreover, the physical
properties of sources associated with IRDCs in our sample are also
consistent with those reported in previous IRDC studies, e.g. Carey
et al. (1998, 2000) with the column density as high as $10^{23}$
cm$^{-2}$ and gas density $\sim10^5-10^6$ cm$^{-3}$. These
physical properties are consistent with the
speculation of Cyganowski et al. (2008) that EGOs are MYSOs.

However, the physical parameters derived here may not be
very reliable as the assumption of the uniform ambient temperature
for all sources and the derived sizes under this assumption may
not reflect the actual temperatures and molecular cloud sizes
for the observed sources. Especially when other sources in addition to
EGOs are present within the telescope beam, the emission from other sources
will contribute to the derived physical parameters e.g. sizes and
masses of EGOs. Thus the derived physical parameters should be treated as
upper limits.

\subsection{Accretion Properties of Infall Candidates}

We can measure an infall velocity (V$_{in}$) using the two layer radiative
transfer model of Myers et al. (1996; Eq. (9) of their work) for each of the 20 sources with blue,
double-peaked HCO$^{+}$ profiles identified in \S 3.2.
Then we can use Eq. (3) of Klaassen \& Wilson (2007)
to calculate the mass infall rate \.{M}$_{in}$ for each of them. The infall
velocity and mass infall rate are summarized in Table 9. In this calculation,
the C$^{18}$O line width (column (14) of Table 6) was adopted as a measure
of the velocity dispersion in the circumstellar material, and the ambient
density n(H$_{2}$) is assumed to be the gas density of cloud
determined from CO line listed in column (11) of Table 7. From our rms
uncertainties in temperatures and one-half the spectral resolution of our observations,
we determined the uncertainties of the infall velocity and infall rate for
each source. From this analysis, we determined the infall velocities
ranging from 0.5 to 8 km s$^{-1}$ with a mean value of 2 km s$^{-1}$,
and the mass infall rates ranging from $4\times10^{-2}$ to $1\times10^{-4}$
M$_{\odot}$ yr$^{-1}$. The infall velocities obtained in our observations
are somewhat higher than those presented in Fuller et al. (2005) and
Klaassen \& Wilson (2007, 2008) with a typical value of $<$1.5 km s$^{-1}$.
This could be due to our larger telescope beam that might contain
more outer material of the cloud with large LOS velocity to produce somehow
larger observable infall velocity. Thus, the calculated infall
velocities may be treated as the upper limit in our observations.
The infall rates derived here are higher than those observed for low-mass
star forming regions, but are consistent with that determined in other
surveys of MYSOs (Fuller et al. 2005; Klassen \& Wilson 2007, 2008) and the theoretical
accretion rates for massive star forming cores (McKee \& Tan 2003;
Bonnell \& Bate 2006).

Usually the mass outflow rates are also orders of magnitude higher in high-mass star
forming regions than that in low-mass star forming regions (e.g. Beuther et al.
2002). Fuller et al. (2005) suggested that the infall rates are consistent
with the outflow rates in their survey, but Klaassen \& Wilson (2007) found
that the ratio of the mass outflow rate to infall rate to be \.{M}$_{in}$/\.{M}$_{out}
$$\approx$1/16 for the UC H{\sc ii} source  Cep A. In our survey,
the mass outflow rates for only two of infall double-peaked profile sources
(G16.59-0.05, G35.20-0.74) are presented by Wu et al. (2004) catalog of
high-velocity outflows. We find the ratio of \.{M}$_{in}$/\.{M}$_{out}$ to
be $2-3$ for the two sources, much higher than that reported by Klaassen \&
Wilson (2007). Behrend \& Maeder (2001) have suggested that the
\.{M}$_{in}$/\.{M}$_{out}$ ratio is probably nonconstant during the
evolution of accreting stars, and could be a decreasing function of the
stellar mass. Combining these factors, we suggest that the EGOs in our survey
should be at very earlier evolutionary stage of massive star formation
and the centre protostellar mass could rapidly increase with a higher
ratio value of \.{M}$_{in}$/\.{M}$_{out}$ compared to the case of UC H{\sc ii}
Cep A at late evolutionary stage reported by Klaassen \& Wilson (2007).
This property is also consistent with the inference of Cyganowski
et al. (2008) that EGOs trace the active rapid accretion earlier
stage of massive protostellar evolution. However the ratio value
of \.{M}$_{in}$/\.{M}$_{out}$ obtained in our observations may be
overestimated due to the large beam.

\subsection{The Limitations for Single-Pointing-Only Surveys}

In this section, we give some discussions or summarizations for the
limitations to our single-pointing-only survey. We can only detect the
possible infall and outflow dynamics from the spectral profile (blue
profile and line wing) with single-pointing observations.
But the actual emission regions of infall and outflow could not be
determined, thus we could not
give more information to understand the dynamics properties of EGOs.
We could not measure the actual observed mm line emitting regions,
although these regions could be simply estimated by the assumption of the uniform
ambient temperature for all sources. But these estimated regions might not
trace the true sizes of molecular clouds around EGOs as discussed in \S 4.2.
Thus the derived physical parameters (i.e. gas density and mass) of EGO
and infall velocity and rate of infall candidates might not reflect
the true cases for EGOs. Especially, the large beam in our observations
would bring large limitations for the studies. We could not
exclude absolutely that the observed emission might be dominated by other
sources in addition to EGOs within the large beam.
The HCO$^{+}$ line profiles could be affected
if there are multiple sources all of which contribute to the emission within
the beam. We also can not determine whether there are multi-core or only one core
to form massive star within EGO regions. We can only say that
the derived physical properties actually trace a relatively large regions (or clumps) therein EGOs embedded.
Moreover our single-pointing-only survey with a large beam still
can not answer where and how the infall dynamics has happened, in an
isolated core (monolithic collapse) or in a protocluster (competitive accretion).
Though we have determined a significant blue excess suggesting infall
signature in full observed EGO sample, some very low or negative blue excess
have also been detected in some particular populations, e.g. IRDC, and
class II methanol maser sub-samples. We could not obviate the possibility that
the competitive accretion might have been occurred in such populations
possibly associated with cluster environments, because the competitive accretion
theory also suggests that the infall may be not statistically significant due to larger velocity
dispersion in complex cluster environments (Bonnell \& Bate 2006).
Future high-resolution observations are very important for detecting the
accretion scale to investigate which is the dominant accretion dynamics in EGOs.

\section{Conclusion}

Using the PMO-13.7 m radio telescope, we performed the first
systematic molecular line (including HCO$^{+}$ and CO lines at 3 mm band)
survey toward a new MYSO sample of 88 EGOs in the northern hemisphere
identified from Spitzer GLIMPSE survey to search for
infall evidence and understand the physical properties in these
sources. We detected HCO$^{+}$ emission in 72 sources.
By analyzing the line profiles of the optically thick
line HCO$^{+}$ and the optically thin line C$^{18}$O for 69
of 72 sources, we identified 29 blue profile candidates with 20 double-peaked
profiles and 9 skewed profiles and, 19 red profile candidates
with 9 double-peaked profiles and 10 skewed profiles. Thus a blue
profile excess of about 0.14 was measured, suggesting that the
infall is statistically significant in our EGO sample. This
value is somewhat different from other surveys towards MYSOs
selected from different criteria, e.g. 6.7 GHz class II methanol masers or
UC H{\sc ii} regions. The blue excess analysis in different sub-samples
was then performed to understand the different blue excess values among
our survey and other surveys. We found that the sources not associated
with IRDCs show a higher blue excess (0.41) than those associated
with IRDCs (-0.08), the ``possible'' outflow candidates also
show a higher blue excess (0.29) than ``likely'' outflow candidates
(0.05), and the blue excess (0.19) found in UC H{\sc ii} regions
is higher than that in 6.7 GHz class II methanol masers (0.07).
These statistical results suggest that a relatively small
blue excess determined in our observations is mainly because that
most observed EGOs are dominated by outflows and are at an early
evolutionary stage associated with IRDCs and 6.7 GHz methanol
masers. By combining the statistical results from our survey
and with those from other surveys, we point out that the infall signatures
gradually evolving with the different stage may be a genuine
property of massive star formation: the blue excess is not statistically
significant in the earlier evolutionary stage
associated with IRDCs or 6.7 GHz class II methanol masers,
and the blue excess will gradually become to be statistically significant
with massive star formation evolution, especially when arriving at
the later evolutionary stage associated with UC H{\sc ii} region.

Furthermore, the physical properties of EGOs are derived using CO lines.
Typical size of the cloud surrounding EGO is 0.8 pc, typical column density is
$2\times10^{23}$ cm$^{-2}$, typical
volume density is about $2\times10^{5}$ cm$^{-3}$, and typical
mass is about $2\times10^3$ M$_{\odot}$, all of which are similar to
those of massive clump wherein massive star forming cores
associated with EGO possibly embedded. These physical properties support the speculation
of Cyganowski et al. (2008) that EGOs are associated with MYSOs.
The estimated infall velocities (typically 2 km s$^{-1}$) and mass
infall rates ($4\times10^{-2}$ to $1\times10^{-4}$ M$_{\odot}$ yr$^{-1}$) for
20 sources with blue double-peaked profiles are also consistent with
that determined in other surveys for MSFRs and the theoretical values.
The higher ratio values of \.{M}$_{in}$/\.{M}$_{out}$$\sim$($2-3$)
measured in two sources (G16.59-0.05, G35.20-0.74) may also suggest
that the EGOs in our survey should be at very earlier evolutionary
stage of massive star formation and the centre protostellar mass could
rapidly increase. However the physical and infall parameters may be
overestimated due to the single-pointing observations with a large beam.

Though there are some limitations due to our single-pointing-only survey with a large
beam for studying the EGOs at present, this survey still offers useful information
to our investigating the properties of EGOs from the statistical view. It
provides a working sample to study the EGOs with high-resolution observations
in future.

 \acknowledgements

We thank an anonymous referee for helpful comments that improved
the manuscript. We are grateful to the staff of Qinghai Station of Purple Mountain
Observatory for their assistance in the observation. This research
has made use of the SIMBAD database, operated at CDS, Strasbourg,
France, and the data products from the GLIMPSE survey, which is a
legacy science program of the {\em Spitzer Space Telescope},
funded by the National Aeronautics and Space Administration. This
work was supported in part by the National Natural Science
Foundation of China (grants 10803017, 10573029, 10621303,
10625314, 10633010, 10673024, 10733030 and 10821302) and the
Knowledge Innovation Program of the Chinese Academy of Sciences
(Grant No. KJCX2-YW-T03), and the National Key Basic Research
Development Program of China (No. 2007CB815405 and 2007CB815403).


\tabletypesize{\scriptsize}

\setlength{\tabcolsep}{0.1in}

\begin{deluxetable}{llccccccccc}
\tablewidth{0pt} \tablecaption{Sample parameters}
\renewcommand{\arraystretch}{0.98}
\tablehead{ &\multicolumn{2}{c}{Position$^a$}& \\
\cline{2-3}
Source & R.A. (2000)& Dec. (2000)& IRDC$^{b}$ & CH$_{3}$OH$^c$ & UC H{\sc ii}$^c$& 1.1 mm$^c$ & Remark$^d$\\
 & ($\emph{h m s}$) & (\arcdeg \ \arcmin \ \arcsec) &  &
}\startdata

   G10.29-0.13 &   18  08  49.3    &   -20 05  57  &   Y   &   Y   &   --  &   Y  &   2  \\
   G10.34-0.14 &   18  09  00.0    &   -20 03  35  &   Y   &   Y   &   --  &   Y  &   2  \\
   G11.11-0.11 &   18  10  28.3    &   -19 22  31  &   Y   &   --  &   --  &   Y  &   3  \\
   G11.92-0.61 &   18  13  58.1    &   -18 54  17  &   Y   &   Y   &   --  &   -- &   1  \\
   G12.02-0.21 &   18  12  40.4    &   -18 37  11  &   Y   &   --  &   --  &   Y  &   1  \\
   G12.20-0.03 &   18  12  23.6    &   -18 22  54  &   N   &   Y   &   --  &   Y  &   4  \\
   G12.42+0.50 &   18  10  51.1    &   -17 55  50  &   N   &   --  &   Y   &   Y  &   4  \\
   G12.68-0.18 &   18  13  54.7    &   -18 01  47  &   N   &   Y   &   --  &   Y  &   4  \\
\enddata

\tablenotetext{a} {The targeted positions in the observations are
from the EGO positions presented by the EGO catalog of Cyganowski
et al. (2008).}

\tablenotetext{b}{Association with IR dark clouds, : Y = Yes, N =
No, presented by Cyganowski et al. (2008).}

\tablenotetext{c}{Associations with 6.7 GHz class II methanol
masers, UC H{\sc ii} regions and 1.1 mm continuum sources within
30\arcsec: Y = Yes, N = No,``--'' = no information, identified from
the 6.7 GHz maser catalogs (Szymczak et al. 2007; Cyganowski et al. 2008, 2009;
Caswell 2009; Xu et al. 2009) and UC H{\sc ii} catalogs
(Wood \& Churchwell 1989; Becker et al. 1994; Kurtz et al. 1994;
Walsh et al. 1998; Forster \& Caswell 2000; Wu et al. 2007),
and 1.1 mm continuum BOLOCAM GPS archive (Rosolowsky et al. 2009).}

\tablenotetext{d}{Remarks: 1, 2, 3, 4 and 5 represent that the
sources are selected from Tables 1, 2, 3, 4 and 5 of Cyganowski et
al. (2008), respectively. And 1, 2 and 5 are classified into
''likely'' outflow candidates, and 3, 4 are classified into
``possible'' outflow candidates.}

\end{deluxetable}

\clearpage


\begin{deluxetable}{lrcccccc}
\tabletypesize{\normalsize}
 \setlength{\tabcolsep}{0.10in}

\tablewidth{0pt} \tablecaption{List of observed lines}

\tablehead{Line & Date & $\nu$ & $\theta_{mb}$ & $\eta_{mb}$ & $\Delta v$\\
  & (UT) & (GHz) & (arcsec) & & (km s$^{-1}$) \\
  (1) & (2) & (3) & (4) & (5) & (6)}
\startdata

HCO$^{+}$ (1-0) & 2009 Jan 9-23  & 89.188518  & 80.6 & 0.50 & 0.205 \\
$^{12}$CO (1-0) & 2009 Feb 12-17 & 115.271202 & 62.4 & 0.61 & 0.369 \\
$^{13}$CO (1-0) & 2009 Feb 12-17 & 110.201353 & 65.3 & 0.60 & 0.114 \\
C$^{18}$O (1-0) & 2009 Feb 12-17 & 109.782173 & 65.5 & 0.60 & 0.115 \\

\enddata

\tablecomments{Col. (1): observed line. Col. (2): observational date.
Col. (3): rest frequency. Col. (4): beam size. Col.
(5): main beam efficiency. Col. (6): spectral resolution.}

\end{deluxetable}


\tabletypesize{\scriptsize}

\setlength{\tabcolsep}{0.08in}


\begin{deluxetable}{lcccccccccccc}
\renewcommand{\arraystretch}{0.93}

\tablewidth{0pt} \tablecaption{Observed HCO$^{+}$ (1--0) line properties of all 88 sample sources.}
\tablehead{ & \multicolumn{4}{c}{Gaussian fits} & & \multicolumn{2}{c}{Peak positions $^a$} \\
\cline{2-5} \cline{7-8}
Source &  $\int$T$_{MB}dV$ & V$_{LSR}$ &  $\Delta V$ & T$_{MB}$ &  & T$_{MB}$ & V$_{LSR}$ & $\sigma_{rms}$  \\
 & (K km s$^{-1}$) & (km s$^{-1}$) & (km s$^{-1}$) & (K) & & (K) & (km s$^{-1}$) & (K) \\
 (1) & (2) & (3) & (4) & (5) & & (6) & (7) & (8)}

\startdata

G10.29-0.13$^b$   &    8.43(0.21)   &    13.39(0.07)   &    5.66(0.17)    &    1.40    &     &   ...   &  ...    &  0.12  \\
G10.34-0.14       &    9.20(0.50)   &    13.22(0.05)   &    3.87(0.16)    &    2.24    &     &   2.31  &  13.31  &  0.16  \\
G11.11-0.11       &    5.18(0.52)   &    29.37(0.27)   &    5.96(0.48)    &    0.81    &     &   0.74  &  27.86  &  0.14  \\
                  &    ...          &    ...           &    ...           &    ...     &     &   0.31  &  33.58  &  ...   \\
G11.92-0.61       &   10.98(0.76)   &    37.66(0.10)   &    6.52(0.29)    &    1.58    &     &   1.08  &  34.25  &  0.16  \\
                  &    ...          &    ...           &    ...           &    ...     &     &   1.20  &  40.00  &  ...   \\
G12.02-0.21$^b$   &    4.34(0.18)   &    -3.65(0.08)   &    3.99(0.20)    &    1.02    &     &   ...   &  ...    &  0.12  \\
G12.20-0.03$^b$   &    5.33(0.17)   &    51.09(0.06)   &    3.96(0.16)    &    1.26    &     &   ...   &  ...    &  0.12  \\
\enddata

\tablecomments{Col. (1): source name. Cols. (2)-(5): the integrated intensity $\int$T$_{MB}dV$,
the velocity at peak V$_{LSR}$, the line width $\Delta$V, and the peak main beam
temperature T$_{MB}$, estimated from Gaussian fits to HCO$^{+}$ lines after masking the absorption
dips and subtracting Gaussian fits to any line wings (see Table 5), the error is given in parenthesis.
Cols. (6)-(7): the real observed T$_{MB}$ and V$_{LSR}$ at HCO$^{+}$
spectral peaks for double-peaked and skewed profiles. Col. (8): 1$\sigma$ noise in the observed HCO$^{+}$
spectrum at T$_{MB}$ scale. ``---'': represents no parameters derived in the sources
without HCO$^{+}$ emission detected; ``...'': represents blank or omitted information.}

\tablenotetext{a}{The peak position is used to analyze line asymmetry (see \S3.2). For non-asymmetric profile
source, the real observed T$_{MB}$ and V$_{LSR}$ adopt the Gaussian fit values in col. (5) and col. (3).
Their uncertainties are assumed to be the 1$\sigma$ noise of HCO$^{+}$ spectrum listed in col. (8) and one-half the
channel width, respectively.}
\tablenotetext{b}{Sources with non-asymmetric profiles (21 in total).}
\tablenotetext{c}{Sources with complex HCO$^{+}$ spectral profiles. They were not used in analysis for the line
asymmetry (see \S3.2).}
\tablenotetext{d} {No HCO$^{+}$ emission detected (16 sources in total).}

\end{deluxetable}


\tabletypesize{\scriptsize}

\setlength{\tabcolsep}{0.10in}


\begin{deluxetable}{lccccc}
\renewcommand{\arraystretch}{1.0}

\tablewidth{0pt} \tablecaption{Gaussian fits to HCO$^{+}$ line wings.}
\tablehead{Source $^a$ &  $\int$T$_{MB}dV$ & V$_{LSR}$ &  $\Delta V$ & T$_{MB}$ & Remark$^b$ \\
 & (K km s$^{-1}$) & (km s$^{-1}$) & (km s$^{-1}$) & (K)  \\
 (1) & (2) & (3) & (4) & (5) & (6) }

\startdata

G10.34-0.14      &    5.04(1.01)   &    11.84(0.87)   &    17.18(3.55)    &    0.28    &  D   \\
G11.92-0.61      &    7.13(1.43)   &    35.71(0.77)   &    22.01(4.33)    &    0.30    &  D   \\
G12.42+0.50      &    3.71(1.01)   &    17.98(0.46)   &     8.93(2.08)    &    0.39    &  D   \\
G12.91-0.26      &   11.21(1.10)   &    38.66(0.24)   &    13.44(0.28)    &    0.79    &  D   \\
G16.58-0.08      &    0.90(0.33)   &    36.53(2.16)   &     5.40(2.28)    &    0.16    &  B   \\
G18.67+0.03      &    2.35(1.11)   &    82.07(1.49)   &    10.89(2.82)    &    0.20    &  R   \\
G18.89-0.47      &    8.17(1.22)   &    66.88(0.49)   &     8.46(2.47)    &    0.90    &  D   \\
G19.01-0.03      &    5.29(0.68)   &    59.50(0.28)   &     6.09(0.70)    &    0.82    &  D   \\
\enddata

\tablecomments{Col. (1): source name. Cols. (2)-(5): the integrated intensity $\int$T$_{MB}dV$,
the velocity at peak V$_{LSR}$, the line width $\Delta$V, and the peak main beam
temperature T$_{MB}$ determined from single broad Gaussian fit to HCO$^{+}$ line wing,
the error is given in parenthesis. Col. (6): remark for line wing shape.}
\tablenotetext{a}{The Gaussian fits listed here were subtracted from the HCO$^{+}$ spectra
before analyzing the main line profile.}
\tablenotetext{b}{``D'' means the line wing emission from both blue- and red-shifted (double)
wings; ``B'' and ``R'' denote the line wing emission mainly from blue-shifted wing and red-shifted
wing, respectively.}
\tablenotetext{c}{As there are strong emissions at the blue- and red-shifted wings
of this source (see Figure 1), two separate components (rather than a single broad component)
were used to fit the strong blue- and red-shifted wings.}

\end{deluxetable}


\begin{deluxetable}{lrcccccccccccccccc}
\tabletypesize{\scriptsize}
 \setlength{\tabcolsep}{0.07in}

\tablewidth{0pt} \tablecaption{Comparison of HCO$^{+}$ (1--0) parameters to Purcell et al. (2006) survey.}
\tablehead{ & \multicolumn{4}{c}{This paper} & & \multicolumn{4}{c}{Purcell et al. (2006)} \\
Antenna diameter: & \multicolumn{4}{c}{14m} & & \multicolumn{4}{c}{22m} \\
Beam size: &  \multicolumn{4}{c}{80 arcsec} & & \multicolumn{4}{c}{35 arcsec}\\
\cline{2-5} \cline{7-10}
Line parameters$^a$: & $\int$T$_{MB}dV$ & V$_{LSR}$ &  $\Delta V$ & T$_{MB}$ & &  $\int$T$_{MB}dV$ & V$_{LSR}$ &  $\Delta V$ & T$_{MB}$ \\
 & (K km s$^{-1}$) & (km s$^{-1}$) & (km s$^{-1}$) & (K) & & (K km s$^{-1}$) & (km s$^{-1}$) & (km s$^{-1}$) & (K)}
\startdata
main line:  \\
           G10.29-0.13  &  8.43 & 13.39 & 5.66 & 1.40 &  & 13.66  &  12.04  &  4.53  &  2.60  \\
           G10.34-0.14  &  9.20 & 13.22 & 3.87 & 2.24 &  & 26.98  &  12.30  &  4.02  &  6.00  \\
           G12.68-0.18  &  3.38 & 58.46 & 3.29 & 0.92 &  &  6.55  &  58.40  &  3.46  &  1.90  \\
           G12.91-0.26  & 15.47 & 38.76 & 5.19 & 1.67 &  & 43.31  &  38.88  & 11.35  &  3.00  \\
           G16.59-0.05  & 14.10 & 59.51 & 3.09 & 2.82 &  & 29.29  &  59.34  &  3.56  &  6.00  \\
           G19.36-0.03  &  4.55 & 28.09 & 6.37 & 0.65 &  & 10.22  &  27.85  &  5.19  &  1.30  \\
           G19.61-0.12  &  2.91 & 57.37 & 3.74 & 0.73 &  &  9.32  &  57.14  &  4.75  &  1.80  \\
\enddata
\tablenotetext{a}{The Gaussian fit parameters for main lines and line wings,
except T$_{MB}$ in main line profile which is the measured peak temperature value of the observed line.}

\end{deluxetable}
\clearpage


\tabletypesize{\tiny}

\setlength{\tabcolsep}{0.05in}

\begin{landscape}

\begin{deluxetable}{llcccccccccllcccccc}

\tablewidth{0pt} \tablecaption{Observed CO (1--0) line properties of 72
sources.}
\tablehead{ & \multicolumn{5}{c}{$^{12}$CO} & & \multicolumn{5}{c}{$^{13}$CO} & & \multicolumn{5}{c}{C$^{18}$O} \\
\cline{2-6} \cline{8-12} \cline{14-18}
Source & $\int$T$_{MB}dV$ & V$_{LSR}$ &  $\Delta V$ & T$_{MB}$ & $\sigma_{rms}$ & & $\int$T$_{MB}dV$ & V$_{LSR}$ &  $\Delta V$ & T$_{MB}$ & $\sigma_{rms}$  & & $\int$T$_{MB}dV$ & V$_{LSR}$ &  $\Delta V$ & T$_{MB}$ & $\sigma_{rms}$   \\
 & (K km s$^{-1}$) & (km s$^{-1}$) & (km s$^{-1}$) & (K) & (K) & & (K km s$^{-1}$) & (km s$^{-1}$) & (km s$^{-1}$) & (K) & (K) & & (K km s$^{-1}$) & (km s$^{-1}$) & (km s$^{-1}$) & (K) & (K) \\
 (1) & (2) & (3) & (4) & (5) & (6) & & (7) & (8) & (9) & (10) & (11)& & (12) & (13) & (14) & (15) & (16) }

\startdata

G10.29-0.13 &  65.90(6.59) & 11.97(0.16) &  5.78(0.41) & 10.70 & 0.23 &  & 54.20(0.62) & 13.08(0.03) & 5.70(0.08) &  8.95 & 0.18 &  & 11.27(0.22) & 13.33(0.04) & 4.49(0.13) & 2.35 & 0.18 \\
G10.34-0.14 & 138.52(1.34) & 11.77(0.37) &  8.33(0.37) & 15.62 & 0.16 &  & 49.87(0.97) & 11.69(0.04) & 4.58(0.11) & 10.40 & 0.20 &  &  9.43(0.20) & 11.84(0.03) & 3.28(0.08) & 2.70 & 0.18 \\
G11.11-0.11 &  51.05(1.51) & 30.38(0.11) &  7.62(0.23) &  6.30 & 0.26 &  & 24.17(0.83) & 30.20(0.02) & 3.34(0.07) &  7.13 & 0.17 &  &  7.35(0.17) & 29.91(0.03) & 2.73(0.07) & 2.53 & 0.18 \\
G11.92-0.61 & 117.18(1.51) & 36.16(0.08) & 13.85(0.21) &  7.93 & 0.13 &  & 33.38(0.38) & 36.50(0.04) & 5.02(0.08) &  6.80 & 0.20 &  &  8.32(0.20) & 36.44(0.04) & 3.25(0.09) & 2.40 & 0.18 \\
G12.02-0.21 &  36.67(0.56) & -2.82(0.36) &  6.26(0.36) &  5.49 & 0.16 &  & 15.57(0.23) & -3.56(0.02) & 3.20(0.05) &  4.57 & 0.20 &  &  4.97(0.13) & -3.73(0.04) & 2.80(0.09) & 1.63 & 0.15 \\
G12.20-0.03 &  60.54(2.85) & 49.72(0.16) &  7.19(0.39) &  7.90 & 0.11 &  & 28.72(0.42) & 50.33(0.04) & 5.40(0.09) &  5.00 & 0.20 &  &  5.50(0.17) & 50.92(0.07) & 4.47(0.17) & 1.18 & 0.13 \\
G12.42+0.50 &  72.16(1.05) & 18.16(0.04) &  5.92(0.10) & 11.44 & 0.21 &  & 38.78(0.20) & 17.93(0.01) & 3.30(0.02) & 11.07 & 0.22 &  &  7.02(0.17) & 18.00(0.03) & 2.58(0.07) & 2.63 & 0.18 \\
G12.68-0.18$^a$ & 103.17(2.77) & 35.25(0.08) &  6.24(0.18) & 15.56 & 0.16 &  & 36.32(0.43) & 35.05(0.02) & 4.37(0.06) &  7.83 & 0.20 &  &  4.25(0.13) &           35.03(0.04) & 3.13(0.12) & 1.28 & 0.18 \\
\enddata

\tablecomments{Col. (1): source name. Cols. (2)-(5): the integrated intensity $\int$T$_{MB}dV$,
the velocity at peak V$_{LSR}$, the line width $\Delta$V, and the peak main beam
temperature T$_{MB}$ for $^{12}$CO line determined from a single-Gaussian fit,
the error is given in parenthesis. Col. (6): 1$\sigma$ noise in the observed $^{12}$CO
spectrum at T$_{MB}$ scale. Cols. (7)-(11): the same as cols. (2)-(6) but for $^{13}$CO;
Cols. (12)-(16): the same as cols. (2)-(6) but for C$^{18}$O.}

\tablenotetext{a}{Sources with complex C$^{18}$O spectral profiles. They were not used in analysis for the line
asymmetry (see \S3.2).}
\tablenotetext{b}{$^{12}$CO was used to calculate the optical depth of C$^{18}$O with typical LTE method for these sources (see \S3.2).}

\end{deluxetable}
\end{landscape}


\tabletypesize{\scriptsize}

\setlength{\tabcolsep}{0.05in}

\begin{deluxetable}{lrcrcrcrrrrrr}

\tablewidth{0pt} \tablecaption{The derived line asymmetric
parameters and source physical properties for 69 sources.}
\tablehead{ & \multicolumn{3}{c}{HCO$^{+}$ Line Asymmetries} & & \multicolumn{7}{c}{Source physical properties}   \\
\cline{2-4} \cline{6-13}
Source & $\delta_{v}$ & \tiny{$T_{MB}(B)/T_{MB}(R)$} & Profile $^a$ & & $\tau_{18}$ & $D$ $^b$ & $f$ & $Size$ & $N$( C$^{18}$O) & $N$(H$_{2}$) & $n$(H$_{2}$)  & $M$\\
  &  &  &  &  &  & (kpc) & & (pc) & ($10^{16}$ cm$^{-2})$ & ($10^{23}$ cm$^{-2}$) & ($10^{5}$ cm$^{-3}$) & (M$_{\odot}$) \\
 (1) & (2) & (3) & (4) &  & (5) & (6) &  (7) & (8) & (9) & (10) & (11)& (12)}

\startdata

G10.29-0.13   &    0.01(0.03)  &    ...         &  N   &      &   0.18      &   2.2  &    0.48  &   0.48   &   3.6   &   2.5   &   2.6  &  1000  \\
G10.34-0.14   &    0.45(0.04)  &    ...         &  R   &      &   0.17      &   2.0  &    0.58  &   0.48   &   2.5   &   1.8   &   1.8  &   720  \\
G11.11-0.11   &   -0.75(0.05)  &    2.39(1.17)  &  B   &      &   0.37      &   3.6  &    0.27  &   0.60   &   4.6   &   3.2   &   2.6  &  2000  \\
G11.92-0.61   &    1.10(0.04)  &    0.90(0.08)  &  R   &      &   0.36      &   3.9  &    0.26  &   0.64   &   5.3   &   3.7   &   2.8  &  2600  \\
G12.02-0.21   &    0.03(0.04)  &    ...         &  N   &      &   0.37      &  17.0  &    0.18  &   2.27   &   4.7   &   3.3   &   0.7  & 29000  \\
G12.20-0.03   &    0.04(0.03)  &    ...         &  N   &      &   0.13      &   4.6  &    0.32  &   0.83   &   2.6   &   1.8   &   1.1  &  2200  \\
G12.42+0.50   &   -0.48(0.05)  &    ...         &  B   &      &   0.13      &   2.4  &    0.72  &   0.65   &   1.5   &   1.1   &   0.8  &   780  \\
G12.91-0.03   &    0.69(0.05)  &    ...         &  R   &      &   0.30      &   4.7  &    0.23  &   0.72   &   3.9   &   2.7   &   1.8  &  2500  \\
\enddata

\tablecomments{Col. (1): source name. Cols. (2)-(3): the HCO$^{+}$ line asymmetry
parameters determined by $\delta v$ and the T$_{MB}$(B)/T$_{MB}$(R) methods (see \S 3.2),
their uncertainties (given in parenthesis) calculated from the rms uncertainties
in temperatures and one-half the channel widths. Col. (4): the line profile. Col. (5):
the optical depth of C$^{18}$O line. Col. (6): the kinematic distance to source. Col. (7):
the beam filling factor. Col. (8): the linear size. Col. (9): the column density of C$^{18}$O.
Col. (10): the column density of H$_{2}$. Col. (11): the volume density of H$_{2}$.
Col. (12): the core mass.}

\tablenotetext{a}{The line profile: ``B'' means
blue profile, ``R'' means red profile, ``N'' means non-asymmetry profile.
We adopt $\delta v$ of $\pm$0.25 as the threshold to define the line asymmetry.
For double-peaked profile, the velocity at the brightest peak
is adopted as the value of $v_{thick}$ to calculate $\delta v$, and we consider
an asymmetry to be significant if the difference (or sum) between T$_{MB}$(B)/T$_{MB}$(R)
and its uncertainty is still larger (or less) than 1 for blue (or red)
double-peaked profile.}

\tablenotetext{b}{The near kinematic distance was adopted for source with near/far
distance ambiguity. The distances to G49.07-0.33, G49.27-0.32, G49.27-0.34 and G59.79+0.63
are assumed to be 5 kpc.}

\end{deluxetable}

\clearpage


\begin{deluxetable}{lrcccccc}
\tabletypesize{\normalsize}
 \setlength{\tabcolsep}{0.20in}

\tablewidth{0pt} \tablecaption{Blue-Excess Statistics}

\tablehead{ & $N_{B}$ & $N_{R}$ & $N_{T}$ & $E$ & $P$ }
 \startdata
\multicolumn{6}{c}{Results from Full Sample}\\ \hline
              & 29 & 19 & 69 & 0.14 & 0.09 \\
\hline \multicolumn{6}{c}{Results from Sources with Distance $<=$
4 kpc}\\ \hline

 & 12 & 10 & 31 & 0.06 & 0.58 \\

\hline

\multicolumn{6}{c}{Results with Respect to Outflow Properties}\\
\hline

likely &  14 & 12 & 41 & 0.05 & 0.42 \\
possible & 15 & 7 & 28 & 0.29 & 0.07 \\

\hline

\multicolumn{6}{c}{Results with Respect to Source Properties}\\
\hline
IRDC & 12 & 15 & 37 & -0.08 & 0.65 \\
Non-IRDC & 17 & 4 & 32 & 0.41 & 0.004 \\
UC H{\sc ii}  & 9 & 6 & 16 & 0.19 & 0.30  \\
CH$_{3}$OH maser & 9 & 7 & 25 & 0.07 & 0.40\\

\enddata
\end{deluxetable}
\clearpage

\begin{deluxetable}{lcr}
\tabletypesize{\scriptsize}
 \setlength{\tabcolsep}{0.10in}

\tablewidth{0pt} \tablecaption{Infall Velocity and Mass Infall Rate.}
\tablehead{ & V$_{in}$$^a$ & \.{M}$_{in}$$^a$  \\
Source & (km s$^{-1}$) & (10$^{-4}$ M$_{\odot}$ yr$^{-1}$)}
\startdata

G11.11-0.11      &    1.3(0.8)       &     43(27)     \\
G12.91-0.26      &    2.9(0.8)       &    394(105)     \\
G16.59-0.05      &    2.7(0.8)       &     59(18)     \\
G17.96+0.08      &    0.7(0.5)       &    4.4(3.5)     \\
G21.24+0.19      &    1.9(0.6)       &     28(9)    \\
G23.01-0.41      &    1.2(0.9)       &     91(70)     \\
G25.38-0.15      &    1.6(0.8)       &    110(58)     \\
G28.28-0.36      &    0.8(0.6)       &     19(15)     \\
G29.91-0.81      &    2.5(2.7)       &     74(81)     \\
G34.26+0.15      &    8.4(1.5)       &    444(82)     \\
G34.28+0.18      &    4.6(2.1)       &    277(124)     \\
G34.39+0.22      &    0.7(0.5)       &     14(9)     \\
G35.20-0.74      &    2.0(0.6)       &     26(7)     \\
G36.01-0.20      &    1.3(0.7)       &     31(15)     \\
G39.10+0.49      &    0.5(0.6)       &    1.3(1.5)     \\
G40.28-0.27      &    0.7(0.5)       &     10(7)      \\
G43.04-0.45      &    2.1(1.2)       &     23(12)     \\
G45.50+0.12      &    2.9(1.8)       &     49(30)     \\
G49.42+0.33      &    2.3(1.4)       &     25(15)     \\
G54.11-0.04      &    5.4(2.1)       &     79(30)     \\

\enddata

\tablenotetext{a}{Infall velocities and mass ifall rates for sources
with blue double-peaked HCO$^{+}$ profiles. Their uncertainties (given
in parenthesis) were calculated from the rms uncertainties
in temperatures and one-half the channel widths.}
\end{deluxetable}
\clearpage

\begin{figure*}
\begin{center}
\scalebox{1.2}[1.2]{\includegraphics[120,-10][700,430]{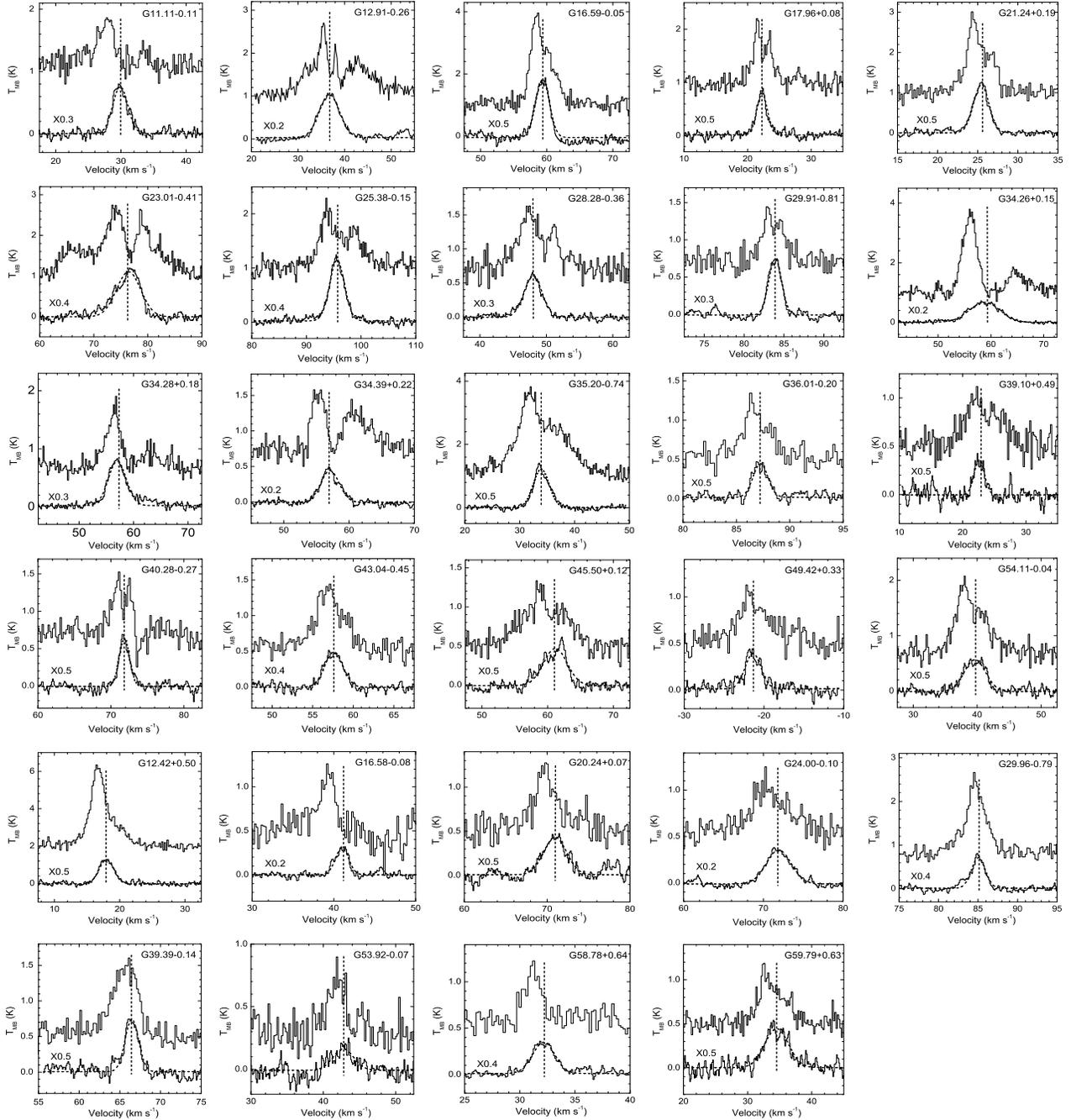}}

\caption{Spectra of HCO$^{+}$ and C$^{18}$O for 29 blue profile
candidates. For each source, the upper spectrum represents the
optically thick line HCO$^{+}$, the lower spectrum represents the
optically thin line C$^{18}$O, the dotted-profile is the Gaussian
fit to the C$^{18}$O line, and the fitted line center velocity is
marked by the dotted-vertical line. The upper four rows
show 20 sources with double-peaked spectra, while the lower two rows show
9 sources with skewed spectra. Note that all the HCO$^{+}$
lines are offset and some (but not all) CO profiles are scaled by
the factors marked above the line.}
\end{center}
\end{figure*}

\begin{figure*}
\scalebox{1.2}[1.2]{\includegraphics[120,140][700,270]{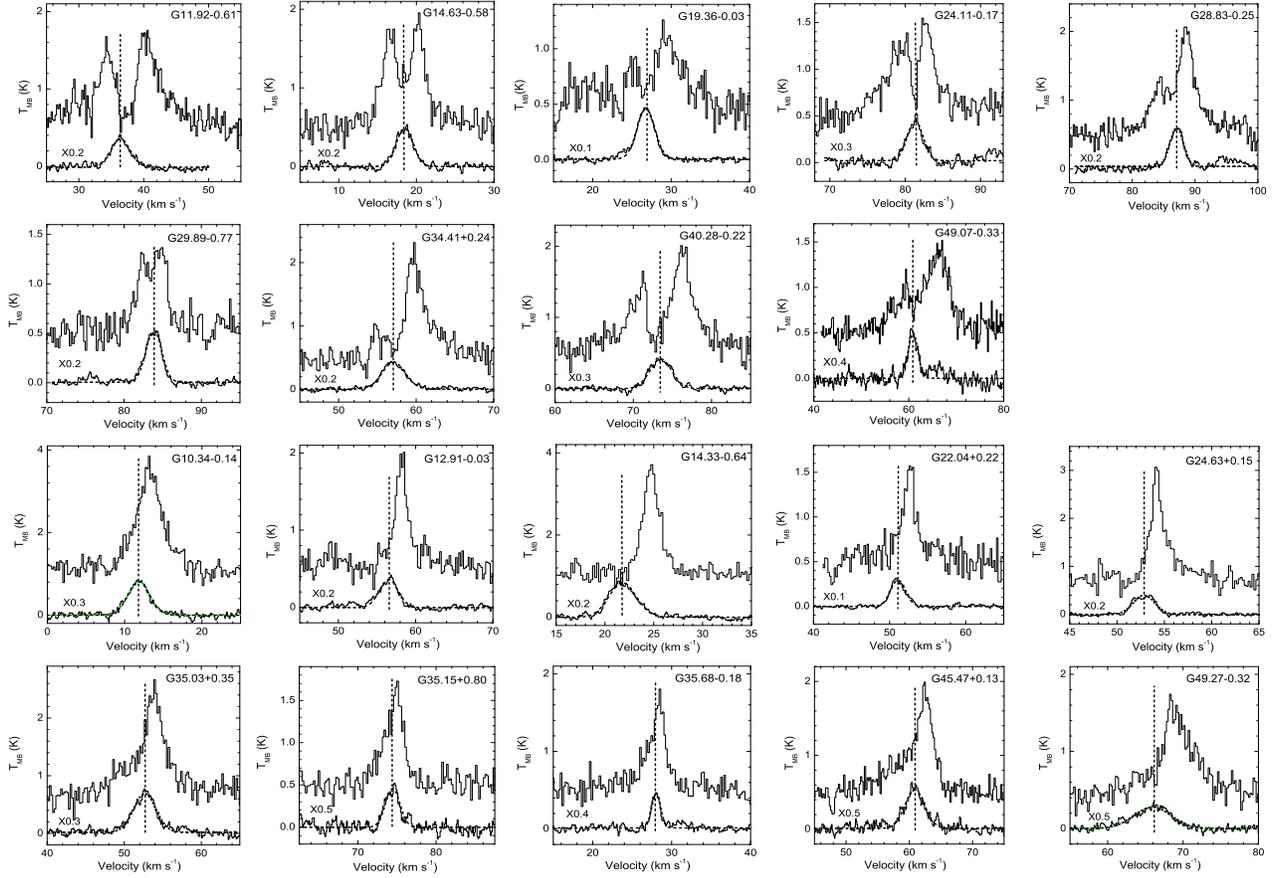}}
\caption{The same as Figure 1 but for 19 red profile candidates.
The upper two rows show 9 sources with double-peaked spectra, and the
lower two rows show 10 sources with skewed spectra.}
\end{figure*}


\begin{figure*}
\begin{center}
\scalebox{1.2}[1.2]{\includegraphics[120,0][700,410]{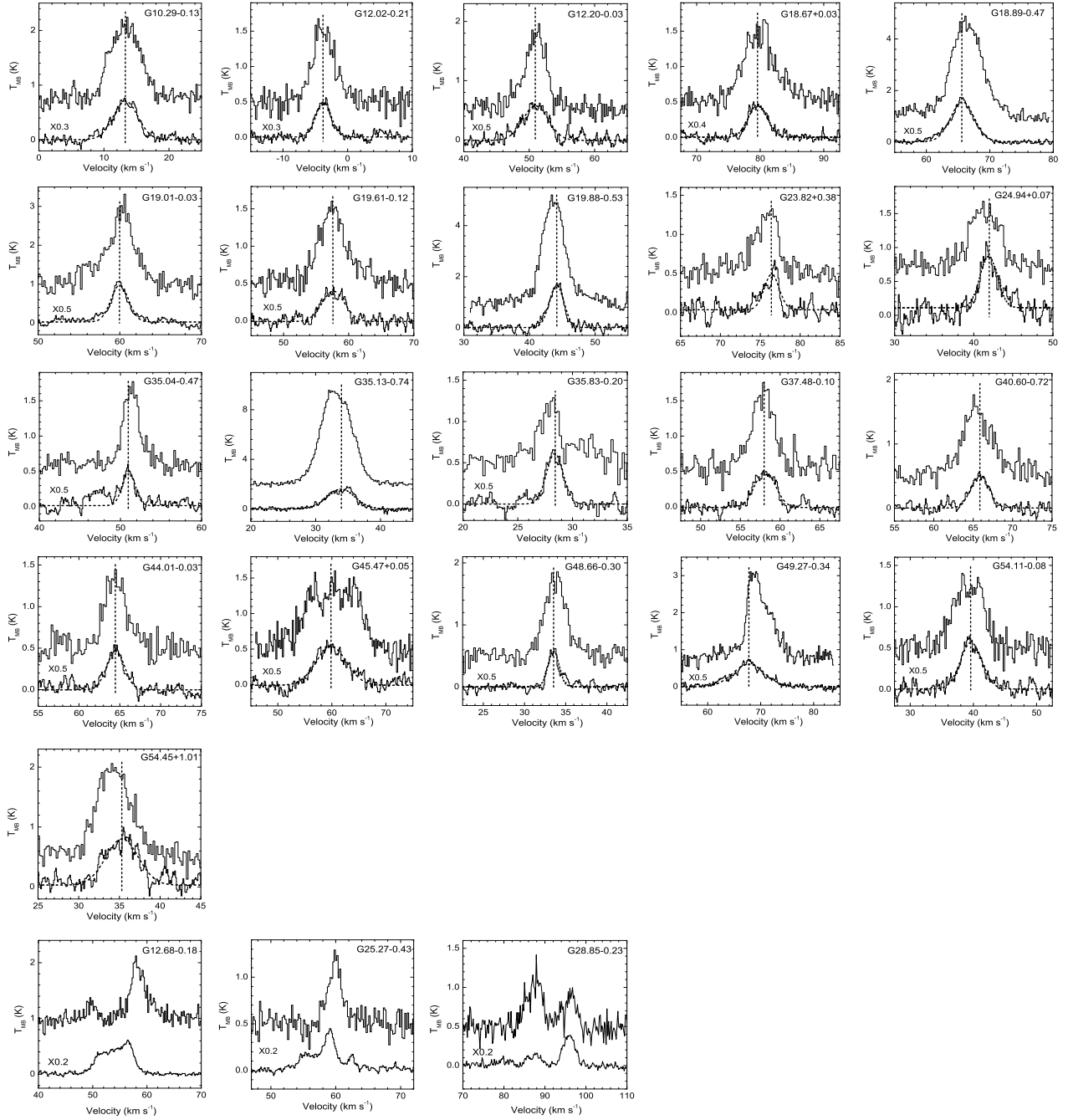}}
\caption{The same as Figure 1 but for 21 sources with non-asymmetric
and 3 sources with complex profiles.
The upper five rows show 21 non-asymmetric profile sources.
The lower row show 3 sources with complex HCO$^{+}$ or CO spectra
which are not used in our line asymmetry analysis (see \S 3.2).}
\end{center}
\end{figure*}


\begin{figure*}[h!]
\begin{center}
\scalebox{0.7}[0.7]{\includegraphics[250,130][400,260]{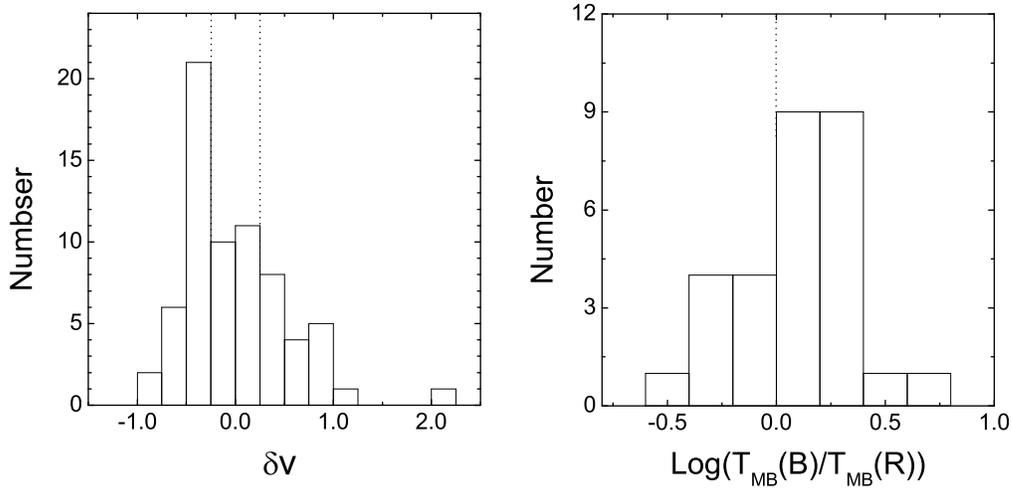}}
\caption{Distribution of $\delta v$ (left; the line asymmetry
parameter; see $\S$ 3.2) and log($T_{MB}(B)$/$T_{MB}(R)$)
(right; ratio of blue vs. red peak intensity). There are 29 blue
profile candidates and 19 red profile candidates among a total of
69 sources identified from $\delta v$ method; and 20 blue profile
candidates and 9 red profile candidates showing double-peaked
spectra identified from brightness ratio method. Note that all
blue or red profiles identified by brightness ratio
method is a subset of the blue or red profiles
identified by the $\delta v$ method (see \S3.2). The vertical dashed
lines in the two panels mark the boundaries for the red or blue profiles.}
\end{center}
\end{figure*}

\end{document}